\documentclass[aps,prb,twocolumn,floatfix,amsmath,amssymb,showpacs,
               superscriptaddress,10pt]{revtex4-1}
\usepackage[final]{graphicx}
\usepackage[caption=false]{subfig}
\usepackage{placeins} 
\usepackage{float}
\usepackage{color}
\usepackage{dcolumn}
\usepackage{xcolor}
\usepackage{url}
\graphicspath{{Figures/}}
\usepackage{mathtools}

\usepackage{bm}
\usepackage{epsfig,psfrag,amsmath,amssymb}
\input{epsf}
\usepackage{hyperref}
\usepackage[percent]{overpic}
\usepackage{physics}

\hypersetup{
    colorlinks=true,
    linkcolor=blue,
    citecolor=blue,
    filecolor=magenta,      
    urlcolor=blue,
}
\usepackage{array}
\usepackage{tabu}
\newcolumntype{C}[1]{>{\centering\arraybackslash$}p{#1}<{$}}

\setcitestyle{square}

\bibpunct{[}{]}{,}{n}{}{}

\usepackage[color=green!60]{todonotes}

\begin{document}
\title{Topological and magnetic properties of the interacting Bernevig-Hughes-Zhang model}

\author{Rahul Soni}
\affiliation{Department of Physics and Astronomy, University of Tennessee, Knoxville, Tennessee 37996, USA}
\affiliation{Institute for Advanced Materials and Manufacturing, University of Tennessee, Knoxville, Tennessee 37996, USA}
\author{Harini Radhakrishnan}
\affiliation{Department of Physics and Astronomy, University of Tennessee, Knoxville, Tennessee 37996, USA}
\affiliation{Institute for Advanced Materials and Manufacturing, University of Tennessee, Knoxville, Tennessee 37996, USA}
\author{Bernd Rosenow}
\affiliation{Institut f{\"u}r Theoretische Physik, Universit{\"a}t Leipzig, Br{\"u}derstrasse 16, 04103 Leipzig, Germany}
\author{Gonzalo Alvarez}
\affiliation{Computational Sciences and Engineering Division, Oak Ridge National Laboratory, Oak Ridge, TN 37831, USA}
\author{Adrian Del Maestro}
\affiliation{Department of Physics and Astronomy, University of Tennessee, Knoxville, Tennessee 37996, USA}
\affiliation{Institute for Advanced Materials and Manufacturing, University of Tennessee, Knoxville, Tennessee 37996, USA}
\affiliation{Min H. Kao Department of Electrical Engineering and Computer Science, University of Tennessee, Knoxville, Tennessee 37996, USA}
\date{\today}

\begin{abstract}
We investigate the effects of electronic correlations on the Bernevig-Hughes-Zhang model using the real-space density matrix renormalization group (DMRG) algorithm. 
We introduce a method to probe topological phase transitions in systems with strong correlations using DMRG, substantiated by an unsupervised machine learning methodology that analyzes the orbital structure of the real-space edges. Including the full multi-orbital Hubbard interaction term, we construct a phase diagram as a function of a gap parameter ($m$) and the Hubbard interaction strength ($U$) via exact DMRG simulations on $N\times 4$ cylinders. Our analysis confirms that the topological phase persists in the presence of interactions, consistent with previous studies, but it also reveals an intriguing phase transition from a paramagnetic to a stripey antiferromagnetic topological insulator.  The combination of the magnetic structure factor, strength of magnetic moments, and the orbitally resolved density, provides real-space information on both topology and magnetism in a strongly correlated system.  
\end{abstract}
\maketitle

\section{Introduction}
Topological insulators are symmetry-protected nontrivial phases of matter featuring conducting edges or surface states while remaining insulating in the bulk \cite{ti-review1,ti-review2,ti-review3}. A prime example is the quantum spin Hall insulator (QSHI) \cite{kane-mele1,bhz-original}, where the underlying time-reversal symmetry (TRS) protects its counter-propagating helical edge modes against non-magnetic impurities \cite{bernevig1,bernevig2,moore06}. Initially proposed for graphene \cite{kane-mele1,kane-mele2}, the QSHI gained attention after its theoretical prediction by Bernevig, Hughes, and Zhang (BHZ) \cite{bhz-original} and subsequent experimental verification by Konig et al. \cite{konig-exp} in two-dimensional (2D) HgTe/CdTe quantum well systems.

While extensively investigated within the framework of non-interacting models \cite{ti-review1,ti-review2}, correlation effects in these systems have drawn much attention recently \cite{int-ti-review}, since the interplay of nontrivial topology and electronic correlations can unveil novel phases of matter. Particularly in the paradigmatic BHZ model, the application of dynamical mean-field theory (DMFT) to investigate correlation effects in the form of inter-orbital and intra-orbital Hubbard interactions has revealed an interaction-induced topological phase transition \cite{tpt-gfn} from a topologically nontrivial phase to a trivial insulator. Another study using inhomogeneous DMFT combined with iterative perturbation theory on BHZ ribbons \cite{tada1} has proposed a topological phase transition from a paramagnetic topological insulating phase to an antiferromagnetic Mott insulating phase. Subsequently, Budich et al. \cite{trauzettel13} provided the first magnetic and topological phase diagram for the interacting BHZ model using DMFT, where they considered a Hund's coupling along with inter- and intra-orbital Hubbard interactions. They found that upon increasing interactions, a non-interacting band insulating state undergoes two phase transitions: identifying a QSH phase at intermediate interactions and a Mott insulating phase in the strongly interacting limit. The transition from the band insulating phase to the QSH phase is of first order \cite{trauzettel15}. The BHZ model with onsite Hubbard-only interaction has also demonstrated the presence of an antiferromagnetic topological insulating (AFTI) phase \cite{topAFM1,topAFM2}, while a DMFT study with strong local Coulomb interactions has postulated the presence of this phase \emph{between} the QSH and the Mott insulating phases \cite{topAFM3}. 

In addition to these interesting interaction effects, TRS-broken BHZ systems have also attracted considerable interest recently, where an in-plane Zeeman term introduced by a ferromagnetic substrate can induce a multitude of topological phenomena such as robust corner states \cite{ren20,chen1}, crystalline Weyl semimetals \cite{dominguez22}, and quantum anomalous Hall effect (QAHE) \cite{saha22}. Additionally, BHZ model with long-range interactions have also demonstrated the presence of QAHE \cite{xue18}. Recent Monte-Carlo studies have also revealed the presence of a topological Mott insulating phase at quarter filling in the generalized spin Hall models \cite{phillips23-1,phillips23-2}. For 3D topological insulators, renormalization group analyses have demonstrated the presence of a topological phase transition into an axionic insulator for strong interactions \cite{bitan16}.  Moreover, the possibility of edge reconstruction in the BHZ model has also been studied in \cite{edge-recon1,edge-recon2,edge-recon3}. Taken together, the diverse and interesting phenomena exhibited by this relatively simple model of a topological insulator combined with advances in the Density Matrix Renormalization Group (DMRG) algorithm as applied to quasi-2D systems \cite{sheng14,hu20,white21,jiang21,kivelson21,jiang23} motivates revisiting with a more exact treatment.

This paper explores the effects of electronic correlations on the BHZ model using a real-space DMRG method on a $N\times 4$ cylindrical geometry. We employ the full multi-orbital Hubbard interaction term involving both  inter- and intra-orbital Hubbard repulsion, Hund's coupling, and a pair hopping term to study correlation effects \cite{kanamori63,patel17,patel20,soni22}.  Using DMRG, we find that while the overall electronic density is unchanged at the cylinder edges, we observe a characteristic increase in $p$-orbital density accompanied by a reduction in $s$-orbital density.  This finding is consistent with large-scale exact diagonalization studies on the corresponding non-interacting model ($U=0$) where the topological phase can be uniquely identify by the presence of zero energy states in the electronic spectrum localized to the edges.  While it is computationally difficult to compute the electronic spectrum in the interacting model with DMRG, orbitally resolved electronic densities are readily available and provide a unique window into the presence of edges states and the accompanying correlated topological phases.  By combining this orbital analysis with local and non-local magnetic properties we construct a phase diagram as a function of gap parameter ($m$) and Hubbard interaction strength ($U$) for $N\times 4$ cylinder (for $N=4,6,8$).  Using orbital densities alone, an unsupervised machine learning approach is sensitive to the existence of two different topologically non-trivial phases, identified by an inversion of the orbital polarity of edge densities.  This distinction is confirmed by an analysis of magnetic properties indicating both a paramagnetic and a topological phase with stripey antiferromagnetic correlations.  The paramagnetic phase was previously identified in DMFT studies \cite{tada1,trauzettel13,trauzettel15}, and here we confirm the presence of a stripey antiferromagnetic topological phase intermediate between the paramagnetic QSH and Mott insulating phases as postulated in \cite{topAFM3}. To aid in the comparison with previous foundational works, we use the same set of parameters considered in \cite{trauzettel13,trauzettel15,topAFM3}.

The organization of the paper is as follows. In Sec. \ref{Sec: Model and Method}, the real-space interacting BHZ model is presented along with the DMRG methodology used in this study. Section \ref{Sec: Results} contains the main numerical results obtained from DMRG calculations of an $N\times 4$ cylinder. Towards the end of this section, we present a complete magnetic and topological phase diagram of the interacting BHZ model using DMRG. Finally, in Sec. \ref{Sec: Conclusion} we conclude and mention a number of implications for future work.

\begin{figure}
\centering
\includegraphics[scale=1.25]{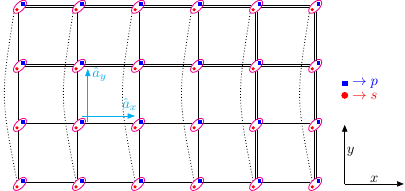}
\caption{Geometry of the $N\times 4$ square lattice cylinder, with $N=6$ unit cells along the $x$ direction and 4 unit cells along the $y$ directions. Each unit cell consists of two orbitals, namely orbital-$s$ and orbital-$p$, marked here as red circles and blue squares, respectively. The cyan arrows indicate the lattice vectors $\hat{a}_x$ and $\hat{a}_y$. The dotted curves are the periodic connections along the $y$ direction. \label{Fig: Lattice Geometry}}
\end{figure}

\section{Model and Method \label{Sec: Model and Method}}
\subsection{Non-interacting BHZ model}
We begin our analysis by examining the real-space Hamiltonian of the BHZ model on a 2D square lattice which is derived from the inverse Fourier transform of the original momentum-space BHZ model \cite{bhz-original}, with a specific focus on the particle-hole symmetric case. The real-space Hamiltonian \cite{trauzettel13,chen1,topAFM1} is given by the equation
\begin{widetext}
\begin{align}
H_{0} &=  B \sum_{\substack{\mathbf{r},\alpha,\sigma}}\left[ c^{\dagger}_{\mathbf{r},\alpha,\sigma}(\tau^z)_{\vphantom{i^i}\alpha\alpha}c_{\vphantom{i^i}\mathbf{r}+\hat{x},\alpha,\sigma} + c^{\dagger}_{\mathbf{r},\alpha,\sigma}(\tau^z)_{\vphantom{i^i}\alpha\alpha}c_{\vphantom{i^i}\mathbf{r}+\hat{y},\alpha,\sigma} + h.c. \right] + m \sum_{\mathbf{r}} (n_{\vphantom{i^i}\mathbf{r},s} - n_{\vphantom{i^i}\mathbf{r},p}) \nonumber \\
& + \frac{A}{2}\sum_{\substack{\mathbf{r},\alpha,\beta,\sigma\\ \alpha\neq\beta}}\left[(-1)^{\sigma}c^{\dagger}_{\mathbf{r},\alpha,\sigma}(\mathfrak{i}\tau^{x})_{\vphantom{i^i}\alpha\beta}c_{\vphantom{i^i}\mathbf{r}+\hat{x},\beta,\sigma} + c^{\dagger}_{\mathbf{r},\alpha,\sigma}(-\mathfrak{i}\tau^{y})_{\vphantom{i^i}\alpha\beta}c_{\vphantom{i^i}\mathbf{r}+\hat{y},\beta,\sigma} + h.c. \right], \label{Eqn: Non-int. Ham}
\end{align} 
\end{widetext}

\noindent
where $\mathbf{r}=(r_{x},r_{y})$ represents the unit cell vector with components $r_{x}$ and $r_{y}$ along the lattice vectors $\hat{a}_x$ and $\hat{a}_y$, respectively, $\alpha$ and $\beta$ denotes the orbital indexes within the unit cell $\mathbf{r}$, where $\alpha,\beta=s,p$ (see Fig.~\ref{Fig: Lattice Geometry} for details), and $\sigma=\uparrow , \downarrow$ represents the $z$-axis spin projection of an electron with orbital $\alpha$ in cell $\mathbf{r}$. The operator $c^{\dagger}_{\mathbf{r},\alpha,\sigma}$ ($c_{\vphantom{i^i}\mathbf{r},\alpha,\sigma}$) creates (annihilates) an electron with spin projection $\sigma$ in orbital $\alpha$ at unit cell $\mathbf{r}$ and $n_{\vphantom{i^i}\mathbf{r},\alpha}=\sum_{\sigma}c^{\dagger}_{\mathbf{r},\alpha,\sigma}c_{\vphantom{i^i}\mathbf{r},\alpha,\sigma}$ is the local density of electrons for orbital $\alpha$ at unit cell $\mathbf{r}$. $\tau^{x}$, $\tau^{y}$, and $\tau^{z}$ are the Pauli matrices. The factor $(-1)^{\sigma}=-1(1)$ for $\sigma=\uparrow(\downarrow)$.

\begin{figure}[!h]
\centering
\hspace*{-0.3cm}\includegraphics[scale=1.4]{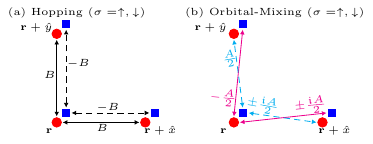}
\caption{(a) Depicts the nearest-neighbor intra-orbital hopping connections along the $x$ and $y$ directions for both spin-up and spin-down particles. The hopping connections are positive for orbital-$s$ and negative for orbital-$p$, respectively. (b) Shows the nearest-neighbor inter-orbital hopping connections or orbital-mixing connections along the $x$ and $y$ directions for both spin-up and spin-down particles. \label{Fig: Hopping Connections}}
\end{figure}

The non-interacting Hamiltonian in Eq.~\eqref{Eqn: Non-int. Ham} is comprised of three separate terms. The first term describes the orbital-dependent nearest-neighbor (NN) intra-orbital hopping with amplitude $B$. Figure~\ref{Fig: Hopping Connections}(a) provides an illustration of these connections, where the hopping amplitude is positive for orbital-$s$ and negative for orbital-$p$, for both spin-up and spin-down particles. The second term is the onsite energy term for the two orbitals, $s$ and $p$. The onsite energy $m$ is positive(negative) for orbital-$s(p)$. The last term accounts for the NN inter-orbital hopping, referred to as the orbital-mixing term within this study. It contains spin-dependent hopping between different orbitals and behaves similar to a NN spin-orbit coupling term. Figure~\ref{Fig: Hopping Connections}(b) demonstrates these connections for spin-up particles along the $x$ and $y$ directions. Despite the presence of this pseudo spin-orbit coupling like term, the single-particle Hamiltonian in equation~\eqref{Eqn: Non-int. Ham} still commutes with the total $S^{z}$ operator, i.e. \@$\left[H_{o},S^{z}_{tot}\right]=0$, making this model a promising candidate to study the interplay of correlation and topology with techniques such as DMRG.

In the remainder of this paper, we have adopted fixed values for the hopping amplitudes: $A=0.3$ and $B=0.5$ to aid comparison with DMFT calculations corresponding to this specific parameter set \cite{trauzettel13,trauzettel15,topAFM3}.

\subsection{Multi-orbital Hubbard interaction}
In order to study the effects of electronic correlations, we consider the more general multi-orbital Hubbard interaction \cite{kanamori63,patel17,patel20,soni22}:
\begin{align}
H_{\rm int} &=  U \sum_{\mathbf{r},\alpha} n_{\vphantom{i^i}\mathbf{r},\alpha,\uparrow}n_{\vphantom{i^i}\mathbf{r},\alpha,\downarrow} + \left(U'-\frac{J_{H}}{2}\right)\sum_{\substack{\mathbf{r} \\ \alpha < \alpha'}}n_{\vphantom{i^i}\mathbf{r},\alpha}n_{\vphantom{i^i}\mathbf{r},\alpha'}  \nonumber \\
&- 2J_{H}\sum_{\substack{\mathbf{r} \\ \alpha < \alpha'}} \mathbf{S}_{\vphantom{i^i}\mathbf{r},\alpha}\cdot\mathbf{S}_{\vphantom{i^i}\mathbf{r},\alpha'} +J_{H}\sum_{\substack{\mathbf{r} \\ \alpha < \alpha'}}\left( P_{\mathbf{r},\alpha}^{\dagger}P_{\vphantom{i^i}\mathbf{r},\alpha'} + h.c \right). \nonumber  \\
\label{Eqn: MOHI}
\end{align}

\noindent
Here, the first term represents the standard on-site Hubbard repulsion $U$ between spin $\uparrow$ and $\downarrow$ electrons, acting on the same orbital within a unit cell. The second term describes the on-site inter-orbital electronic repulsion between electrons at different orbitals within the same unit cell. The third term involves the Hund's coupling $J_{H}$ that explicitly accompanies the ferromagnetic character between the orbitals. The operator $\mathbf{S}_{\vphantom{i^i}\mathbf{r},\alpha}$ denotes the total spin of the orbital $\alpha$ at cell $\mathbf{r}$. The last term signifies the on-site inter-orbital electron-pair hopping $P_{\vphantom{i^i}\mathbf{r},\alpha}=c_{\vphantom{i^i}\mathbf{r},\alpha,\uparrow}c_{\vphantom{i^i}\mathbf{r},\alpha,\downarrow}$.  We incorporate the standard relation $U'=U-2J_{H}$ and fix $J_H/U=0.25$ \cite{trauzettel13,trauzettel15,topAFM3}. The choice $J_H/U=0.25$ facilitates a direct comparison with existing DMFT studies and is believed to be relevant for a wide range of 
transition metal compounds, like the iron-based superconductors \cite{patel17,patel20}.

Equations~\eqref{Eqn: Non-int. Ham} and \eqref{Eqn: MOHI} constitute our interacting BHZ Hamiltonian, given by:
\begin{align}
H &=H_0 + H_{\rm int}.\label{Eqn: Full Ham}
\end{align}
\noindent
To study this Hamiltonian numerically, we performed extensive DMRG \cite{white92,schollwock05} simulations on $4\times 4$ and $6\times 4$ cylinders, that correspond to $16$-site $2$-orbitals and $24$-site $2$-orbitals system, respectively, at half-filling. The featured systems possess open boundary conditions along the $x$-direction and periodic boundary conditions along the $y$-direction. We employ the \texttt{DMRG++} computer program developed by one of the authors (G.A.) \cite{gonzalo09}. To ensure proper convergence of the DMRG calculations, we consider a minimum of 1500 states and a maximum truncation error of $10^{-5}$ throughout the finite algorithm sweeps. By adhering to these criteria, we obtain essentially exact ground-state properties which are used to identify magnetic and topological features and phases of the interacting BHZ model. A particular case of the interacting BHZ model shown in \eqref{Eqn: Full Ham} has also been studied for spinless fermions in \cite{bermudez1,bermudez2}.
 
\section{Results \label{Sec: Results}}
\subsection{Non-interacting $U=0$}
\begin{figure}[!h]
\centering
\hspace{-0.5cm}\includegraphics[scale=0.95]{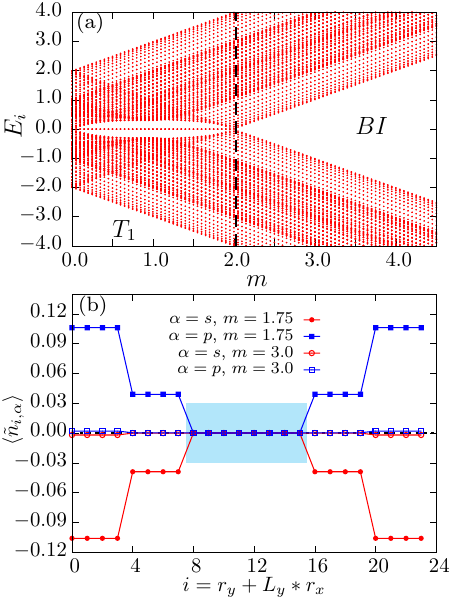}\\
\hspace{0.5cm}\includegraphics[scale=1.1]{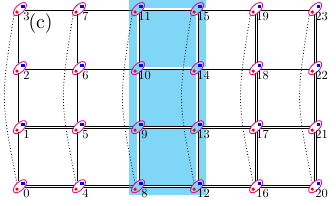}
\caption{(a) Plot of single-particle energy eigenvalues versus the onsite energy $m$ at $U=0$ for a $20\times 4$ cylinder. The figure illustrates the presence of two phases: a non-trivial topological phase T$_1$ and a trivial band insulator BI at $U=0$. (b) Reduced electronic charge density (Eq.~\ref{Eqn: Edge electronic charge density}) at half-filling plotted versus the cell-index ($i$) for $6\times 4$ cylinder at $m=1.75$ and $m=3.0$ within the T$_1$ phase and BI phase, respectively, exhibiting the localization of zero-energy modes on the edges in T$_1$ phase. The cyan box indicates the bulk cells of the cylinder. (c) Lattice geometry of the $6\times 4$ BHZ cylinder with numbered unit cells where the bulk cells are highlighted in cyan.} \label{Fig: Nx4 U=0 Results}
\end{figure}

To aid in the interpretation of interaction effects, we begin by examining the real-space BHZ model at $U=0$ at half-filling on a $N\times 4$ cylinder, evaluating key topological properties via exact diagonalization. In the case of $U=0$, the system clearly shows a topological phase transition from a non-trivial topological phase T$_1$ to a trivial band insulator phase BI, as depicted in Fig.~\ref{Fig: Nx4 U=0 Results}(a) where we plot the single particle energy eigenvalues for a $20\times 4$ cylinder. Here, the topological phase T$_1$ is characterized by the presence of degenerate zero-energy modes for a range of $m\in (0,2]$, a hallmark of the QSHI phase in HgTe/CdTe quantum wells described by the BHZ model \cite{bhz-original}, whereas the band insulator, BI, has a clear gap (with approximate size $2(m-2)$).

To understand how the energy spectrum corresponds to physics at the sample boundary, Figure~\ref{Fig: Nx4 U=0 Results}(b) shows the edge electronic charge density $\langle \tilde{n}_{i,\alpha}\rangle$ of a $6\times 4$ cylinder, which is obtained by subtracting the average charge density of the bulk from the charge density at each cell. More specifically we compute,
\begin{equation}
\langle \tilde{n}_{i,\alpha}\rangle = \langle n_{i,\alpha}\rangle - \frac{1}{N_{b}}\sum_{i\in 
 \rm bulk}\langle n_{i,\alpha}\rangle , \label{Eqn: Edge electronic charge density}
\end{equation}
where $N_{b}$ corresponds to the number of unit cells in the bulk of the cylinder (as highlighted in Fig.~\ref{Fig: Nx4 U=0 Results}(c)) and $i=r_y + L_y*r_x$ is a flattened unit cell index. The edge charge density will act as a topological marker for the cylindrical system in both the interacting and non-interacting case. The fundamental idea is to examine the extent and character of the electronic charge density on the edges in comparison to the bulk. In Fig.~\ref{Fig: Nx4 U=0 Results}(b), we plotted this quantity for two different values of $m$ at $U=0$. The figure clearly illustrates that the edge charge density is quite large for $m=1.75$ (filled symbols) which lies in the topological T$_1$ phase, as compared to $m=3.0$ (open symbols) which lies in the non-topological BI phase. Moreover, these densities are equal and opposite for both orbital-$s$ and orbital-$p$ on either edges of the cylinder. The steps in density observed for $m=1.75$ are indicative of a physical edge states which possess decaying real-space wavefunctions \cite{edge-decay1,edge-decay2,ren20}.  The presence of a finite edge density is indicative of the presence of the zero energy modes, which we expect to be true even in the presence of interactions where the bulk-boundary correspondence still holds \cite{essin11}.

\subsection{Interacting $U>0$}
Having understood the emergent real-space structure of gapless edge states in the non-interacting model, we now discuss DMRG results of the interacting BHZ model on $N\times 4$ cylinders.  

In the subsequent subsections, we present interaction results for both $8\times 4$ and $4\times 4$ cylinders. We begin by demonstrating the consistency of results for both cylinder sizes for a fixed value of the onsite energy $m=3.0$, as a function of increasing $U$. Next, we showcase results for $m=1.75$ in the context of a $4\times 4$ cylinder. Owing to the significant computational cost of $N=8$ and the finite size convergence study presented in Appendix~\ref{Sec: Finite-size scaling}, a majority of results are confined to $N=4$.

\subsubsection{Charge and magnetic properties}
We begin our analysis by focusing on more conventional local and non-local charge and magnetic properties of the system.  We first fix $m=3.0$, deep in the BI phase at $U=0$, and increase $U$ from the weak to strong coupling regime on a $8\times 4$ cylinder. Local charge properties can be quantified through the average occupation of orbitals,
\begin{equation}
\langle n_\alpha \rangle = \frac{1}{N_c}\sum_{\mathbf{r}} \langle n_{\mathbf{r},\alpha} \rangle\, ,
\end{equation}

\begin{figure}[!h]
\centering
\hspace*{-0.5cm}\includegraphics[scale=0.9]{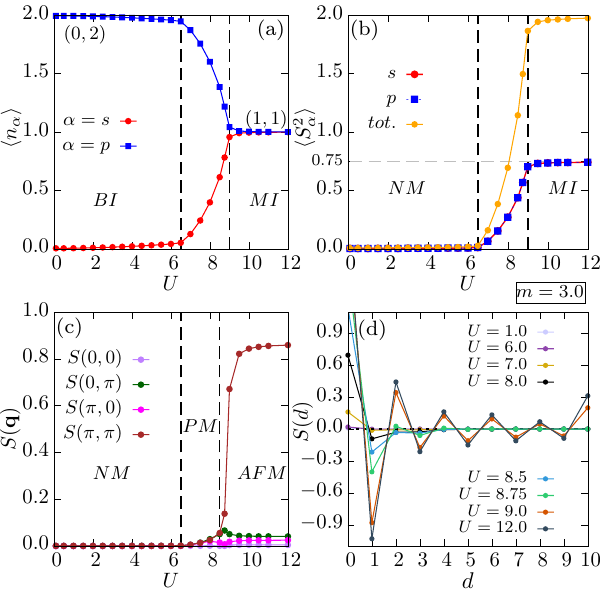}
\caption{Charge and magnetic properties of the interacting BHZ model with $m=3.0$, $U \ge 0$ on a $8\times 4$ cylinder. (a) Average orbital occupation $\langle n_\alpha\rangle$ plotted versus $U$ for both orbitals. The plot clearly depicts a change in orbital occupation from a $(0,2)$ band insulator configuration to $(1,1)$ Mott insulator (MI) configuration as interaction strength is increased. (b) Average magnetic moment of an orbital $\langle S^{2}_\alpha\rangle$ and of the cell $\langle S^{2}_{tot.}\rangle$ plotted versus $U$. The plot shows a magnetic transition from a non-magnetic (NM) phase to a magnetic phase at $U=6.5$. (c) Spin structure factor $S(\mathbf{q})$ showing the transition from NM to PM to AFM phase as we increase $U$. (d) Real-space spin-spin correlations plotted versus the distance between spins.} \label{Fig: Magnetic properties m=3.0, U!=0, 8x4}
\end{figure}

\noindent
where $N_c$ is the total number of unit cells. Figure~\ref{Fig: Magnetic properties m=3.0, U!=0, 8x4}(a) illustrates the  average orbital occupation versus the interaction strength. The plot demonstrates that as we increase $U$ from the weak to the strong coupling regime, the system transitions from a $(n_s,n_p)=(0,2)$ configuration of a band insulator (BI) towards a $(1,1)$ Mott insulator (MI). Between these two insulting phases, beginning near $U=6.5$ we observe a $(n,2-n)$ configuration \cite{werner07} which uniformly changes as we increase the interaction strength until $U=9.0$, where $n$ is the average occupation of orbital-$s$ at interaction strength $U$. This is consistent with previous DMFT results \cite{trauzettel13}, where a similar behavior in the average charge occupation was also observed.

For local magnetic properties, we study the average magnetic moment of the individual orbitals and the average magnetic moment of the entire unit cell,
\begin{align}
\langle S^{2}_\alpha \rangle &= \frac{1}{N_c}\sum_{\mathbf{r}} \langle \mathbf{S}_{\mathbf{r},\alpha}\cdot \mathbf{S}_{\mathbf{r},\alpha} \rangle , \\
\langle S^{2}_{tot.}\rangle &= \frac{1}{N_c}\sum_{\mathbf{r}} \langle \mathbf{S}_{\mathbf{r}}\cdot \mathbf{S}_{\mathbf{r}}  \rangle ,
\end{align}
where $\mathbf{S}_{\mathbf{r}}=\sum_{\alpha}\mathbf{S}_{\mathbf{r},\alpha}$. Figure~\ref{Fig: Magnetic properties m=3.0, U!=0, 8x4}(b) shows the average magnetic moment of the individual orbitals as well as the total for the unit cell as a function of interaction strength at $m=3.0$. The plot shows that the BI phase -- present for $U\leq 6.5$ -- is also non-magnetic (NM), as the average magnetic moments for both orbitals and the unit cell is vanishingly small. For $U>6.5$, the system attains magnetism, and upon increasing interactions  the average magnetic moment of the orbitals smoothly increases until it saturates to $0.75$ in the $(1,1)$ configuration of the Mott insulator for $U>9.0$. Additionally, the magnetic moment for both orbital-$s$ and orbital-$p$ fall directly on top of each other for all values of $U$ in Fig.~\ref{Fig: Magnetic properties m=3.0, U!=0, 8x4}(b); this feature is observed to be true for all values of $m$.

\begin{figure}[t]
\centering
\hspace*{-0.5cm}\includegraphics[scale=0.9]{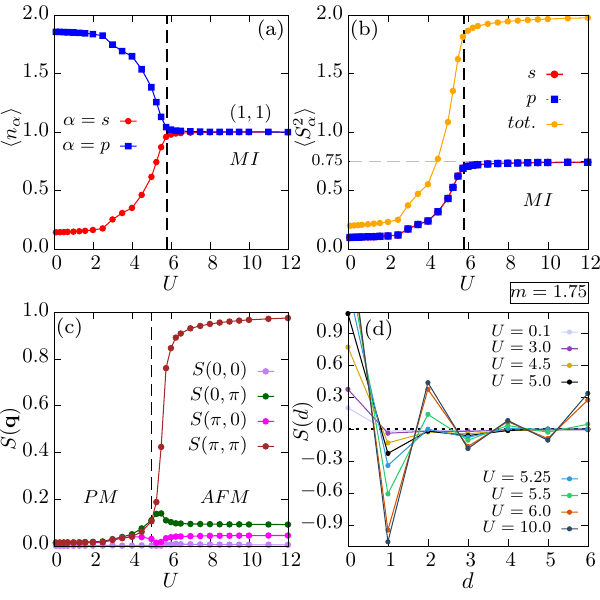}
\caption{Charge and magnetic properties of the interacting BHZ model at $m=1.75$, $U> 0$ on a $4\times 4$ cylinder. (a) Average orbital occupation versus $U$ for both orbital-$s$ and orbital-$p$. The orbital occupation changes towards a $(1,1)$ Mott insulator (MI) configuration as $U$ increases. (b) Average magnetic moment of orbitals and the unit cell as a function of $U$ showing the evolution of the moments from paramagnetic (PM) phase to the MI phase. (c) Spin structure factor $S(\vb{q})$ versus $U$ providing evidence of the magnetic phase transition from PM to an antiferromagnetic Mott insulating (AFM MI) phase as $U$ increases. (d) Real-space spin-spin correlations plotted versus the distance between spins. \label{Fig: Magnetic properties m=1.75, U!=0}}
\end{figure}

Finally, to explore non-local magnetic properties, we compute the real-space spin-spin correlation $S(d)$ defined as:
\begin{equation}
S(d) = \frac{1}{\#(d)}\sum_{i} \langle \mathbf{S}_{i}\cdot \mathbf{S}_{i+d}  \rangle\, ,
\end{equation}
and the corresponding spin structure factor $S(\mathbf{q})$, 
\begin{equation}
S(\mathbf{q}) = \frac{1}{4N_c^2}\sum_{\mathbf{r},\mathbf{r}',\alpha,\beta}e^{-\mathfrak{i}\mathbf{q}\cdot (\mathbf{r}-\mathbf{r}')} \langle \mathbf{S}_{\mathbf{r},\alpha}\cdot \mathbf{S}_{\mathbf{r}',\beta}  \rangle\, ,
\end{equation}
\noindent 
where $\#(d)$ is the number of sites separated by the distance $d$. The structure factor is shown in Fig.~\ref{Fig: Magnetic properties m=3.0, U!=0, 8x4}(c), and it provides a clear picture of the magnetic phases in the system:~(i) it confirms the presence of the NM phase for $U\leq 6.5$,~(ii) for $6.5<U<8.5$ the system is paramagnetic (PM) as there is no dominant magnetic ordering present; and~(iii) for $U\geq 8.5$ the system is antiferromagnetic (AFM) with dominant 
 $(\pi,\pi)$ ordering. Additionally it is worth pointing out that the AFM correlations appears before than the MI phase: AFM ordering is observable at $U=8.5$, while the MI appears for $U>9.0$. Note that the vertical dashed lines in Fig.~\ref{Fig: Magnetic properties m=3.0, U!=0, 8x4}(a-c) serves as an indicator of a qualitative change in observed quantities and does not represent a thermodynamic limit phase boundary.

Figure~\ref{Fig: Magnetic properties m=3.0, U!=0, 8x4}(d) presents a real-space  picture of the magnetic and non-magnetic phases for fixed $m=3.0$. The spin-spin correlations also provides a confirmation of the paramagnetic (PM) phase, where $S(d)$ lacks any signatures of long-range ordering. Additionally, it highlights the onset of the antiferromagnetic (AFM) phase at $U=8.5$. We note that the spin-spin correlations are computed for the entire cells, and we have observed a consistent behavior across individual orbitals for all magnetic and non-magnetic regimes within our calculations.
We note that interaction results, covering both the charge and magnetic properties discussed in this section, as well as the topological properties addressed in the subsequent section for $m=3.0$ in $8\times 4$ cylinders are in alignment with those observed for their $4\times 4$ counterparts. The equivalent results for the smaller cylinders are included in Appendix~\ref{Sec: 4x4 Results at m=3.0} for completeness.

We next present analogous results for a smaller value of $m=1.75$ on a $4\times 4$ cylinder in Fig.~\ref{Fig: Magnetic properties m=1.75, U!=0}, where the non-interacting model lies within the topological T$_1$ phase at $U=0$.  By increasing $U$, we explore the effects of interactions and find that in contrast to the case of $m=3.0$, for $m=1.75$, there is no evidence of a non-magnetic BI phase. This can be observed in panel (a), where the orbital densities are never saturated for small $U$, and instead they continuously move towards the $(1,1)$ occupations characteristic of the Mott insulating phase which develops near $U=6.0$.  The accompanying magnetic moments shown in panel (b) are also distinct from $m=3.0$, where the strictly non-magnetic region is never observed with $\langle S^2_\alpha \rangle > 0$ for all $U\geq 0$. Furthermore, an analysis of the spin structure factor in panel (c) shows that the PM phase persists until nearly $U=5.0$.  However, here we do observe a similarity with $m=3.0$ case, as a weak AFM phase develops, after the PM phase, for $U\geq 5.0$ which continues until it merges into the more robust antiferromagnetic Mott insulating phase at $U=6.0$.  The resulting real-space spin-spin correlations in panel (d) are also larger in the strong interaction limit.  

\begin{figure}[!h]
\centering
\hspace*{-0.5cm}\includegraphics[scale=1]{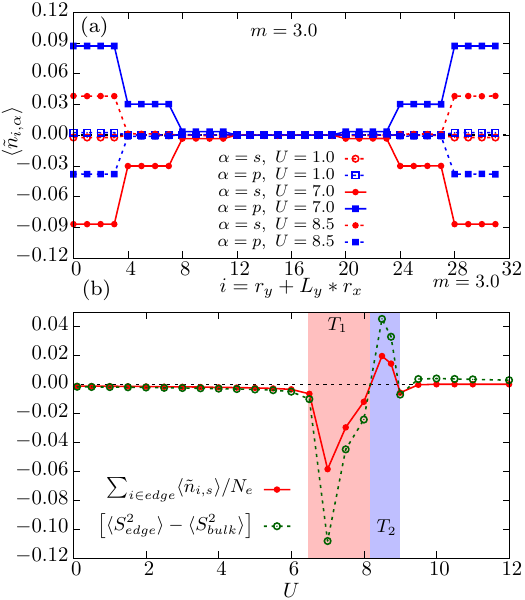}
\caption{Topological properties of the interacting BHZ model at $m=3.0$ for a $8\times 4$ cylinder. (a) Edge electronic charge density versus the cell index $i$ for $U=1.0,~7.0,~8.5$. The plot shows that the parameter point $U=1.0$ is non-topological or trivial, whereas the parameter points $U=7.0~\text{and}~8.5$ are topological. Additionally, the polarity of the edge density for orbital-$s$($p$) are negative(positive) and positive(negative) at $U=7.0~\text{and}~8.5$, respectively, suggesting two different topological phases T$_1$ and T$_2$. (b) Averaged edge density for orbital-$s$ and the difference of edge and bulk magnetic moments plotted versus the interaction strength $U$. The plot shows the change in polarity of the edge density at $U\approx 8.2$ as it moves from the T$_1$ to T$_2$ topological phase. The plot also depicts that the T$_1$ phase is dominated by the bulk magnetic moments, while the T$_2$ phase is dominated by the edge magnetic moments.} \label{Fig: Topological properties m=3.0, U!=0, 8x4}
\end{figure}

\subsubsection{Topological properties}
Motivated by our analysis for the non-interacting BHZ model with $U=0$, we employ the edge electronic charge density defined in Eq.~\eqref{Eqn: Edge electronic charge density} as a key property for the characterization of the topological phases with interaction. In Fig.~\ref{Fig: Topological properties m=3.0, U!=0, 8x4}(a),we illustrate this quantity for three different values of $U=1.0,7.0$ and $8.5$ at $m=3.0$ for a $8\times 4$ cylinder. The interaction $U=1.0$ (open symbols) falls in the band insulating region, as discussed previously. The edge electronic charge density at this parameter value is almost negligible supporting the conclusion that the band insulator is trivial. Conversely, for $U=7.0$ (solid points/lines), the system is in the paramagnetic region and exhibits finite edge density, indicating a non-trivial topology. 
The polarity of the edge-electronic charge density (relative signs of $\expval{\tilde{n}_{i,s}}  < 0 < \expval{\tilde{n}_{i,p}}$ for $i \in \text{edge}$) at $U=7.0$ is consistent with the T$_1$ phase from our non-interacting results. More interestingly, accompanying the onset of antiferromagnetic order near $U=8.5$ (solid points, dashed lines), we observed non-trivial topological behavior in the edge electronic density, but with opposite orbital polarity as compared to the T$_1$ phase at $U=7.0$. i.e. $\expval{\tilde{n}_{i,p}} < 0 < \expval{\tilde{n}_{i,s}}$ for $i \in \text{edge}$.  We use this polarity inversion to infer the existence of a topological T$_2$ phase which exhibits a stripey AFM order as characterized by an asymmetry in the structure factors $S(0,\pi)$ and $S(\pi,0)$.

This inversion is quantified in Fig.~\ref{Fig: Topological properties m=3.0, U!=0, 8x4}(b), where we have computed the edge density for the orbital-$s$, averaged over all the unit cells situated on the edges ($N_e$) of the cylinder and plotted it versus the interaction strength $U$. The result clearly illustrates a distinct polarity change, transitioning from negative to positive, in the edge density of orbital-$s$ as the system magnetically transforms from the paramagnetic region towards the onset of antiferromagnetic order. We take this as evidence indicative of a T$_1$ to T$_2$ topological phase transition.

This change is also accompanied by a difference in magnetic moments of the edge and bulk, which follows the same pattern as the averaged edge electronic charge density. This can be quantified by defining:
\begin{equation}
\Delta S = \langle S^2_{\text{edge}}\rangle- \langle S^2_{\text{bulk}}\rangle
\end{equation}
where 
\begin{align}
    \langle S^2_{\text{edge}}\rangle &= \frac{1}{N_e} \sum_{i\in \text{edge}}\langle S^2_{i}\rangle \\
    \langle S^2_{\text{bulk}}\rangle & = \frac{1}{N_b}\sum_{i\in \text{bulk}}\langle S^2_{i}\rangle\, .
\end{align}
$\Delta S$ is shown as a function of interaction strength in Fig.~\ref{Fig: Topological properties m=3.0, U!=0, 8x4}(b), where we observe that the T$_1$ phase is dominated by the  magnetic moments of the bulk,  in contrast to the T$_2$ phase, which  exhibits larger magnetic moments along the edges.  This can be intuitively understand by appealing to a real-space picture, as the antiferromagnetism first develop on the edges and then moves towards the bulk as $U$ increases, and eventually reach a saturation in the Mott insulating region~\cite{topAFM3}. The presence of the antiferromagnetic ordering and topology has previously been explored at the mean-field level in these systems \cite{topAFM1,topAFM2,topAFM3}. We note that this subtle yet finite feature has a one-to-one connection with the edge electronic density as depicted in the figure.  Similar to the case of charge and magnetic properties for $m=3.0$, we also find a consistency in the topological properties between the $8\times 4$ and $4\times 4$ cylinders depicted in Appendix~\ref{Sec: 4x4 Results at m=3.0}.

\begin{figure}[t]
\centering
\hspace*{-0.5cm}\includegraphics[scale=1]{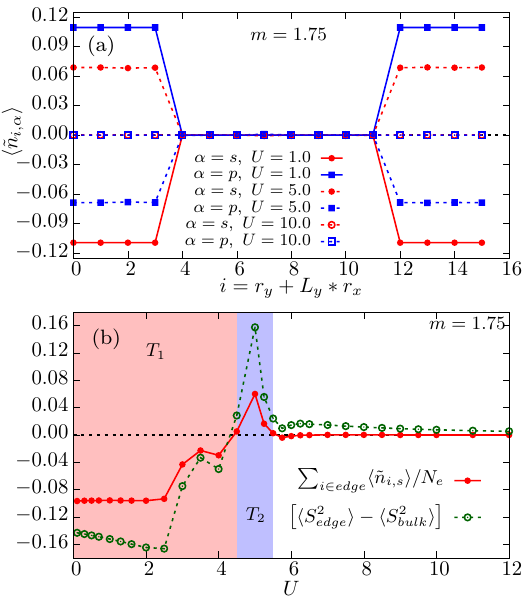} \caption{Topological properties of the interacting BHZ model at $m=1.75$ for a $4\times 4$ cylinder. (a) Edge electronic charge density versus the cell index $i$ for different values of $U(=1.0,~5.0,~10.0)$. The plot illustrates that paramagnetic point $U=1.0$ lies in T$_1$ topological phase, whereas $U=5.0$ lies in the T$_2$ topological phase. Additionally, the point $U=10.0$ which lies in the Mott insulating region is non-topological. (b) Plot of averaged edge density for orbital-$s$ and the difference of edge and bulk magnetic moments versus the interaction strength $U$. The T$_1$ phase covers the entire paramagnetic region and transitions to the T$_2$ phase along with the onset of antiferromagnetic order.  The entire Mott insulating region is observed to be non-topological.  \label{Fig: Topological properties m=1.75, U!=0}} 
\end{figure}

Figure~\ref{Fig: Topological properties m=1.75, U!=0} presents these topological quantities for the case of $m=1.75$ on a $4\times 4$ cylinder. Similar to the observations made for the case of $m=3.0$, we first identify the T$_1$ topological phase within the paramagnetic region, followed by the emergence of T$_2$ topological phase during the onset of antiferromagnetic order. The difference in magnetic moments of the edge and bulk also follows the same behaviour as shown previously for the case of $m=3.0$, that is, the paramagnetic T$_1$ region is dominated by the bulk moments whereas the antiferromagnetic T$_2$ phase is dominated by the edge moments. Moreover, our results also illustrate that the edge electronic charge density in the antiferromagnetic Mott insulating region are negligible, implying that this region is non-topological.

\subsubsection{DMRG phase diagram of the interacting BHZ model}
\begin{figure}[!h]
\centering
\includegraphics[scale=0.98]{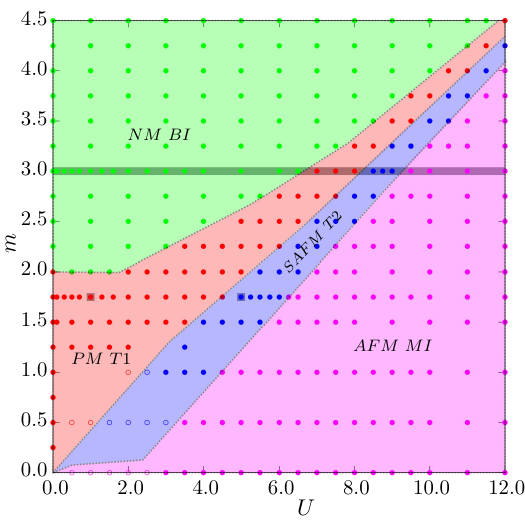}
\caption{Magnetic and topological phase diagram ($m$ vs $U$) of the interacting BHZ model for the $N\times 4$ cylinder, using DMRG. The green region is the non-magnetic trivial band-insulating phase, the red region is the paramagnetic topological T$_1$ phase, the blue region is the topological T$_2$ phase with stripey antiferromagnetic correlations, and the magenta region is the non-topological AFM MI phase. Circles represent simulations for $4\times 4$ cylinders, gray squares and the entire strip at $m=3.0$  were performed for $8\times 4$ cylinders. Solid points indicate DMRG results fully satisfying the convergence criteria detailed in section~\ref{Sec: Model and Method}, while open circles are for parameters where achieving complete convergence of all observables is challenging.} \label{Fig: Magnetic and topological phase diagram}
\end{figure}

Utilizing the analysis presented in the preceding two subsections for the identification of the magnetic and topological phases, we have generated a DMRG phase diagram of the interacting BHZ model.  The result, presented in Fig.~\ref{Fig: Magnetic and topological phase diagram} shows the phases as a function of the gap parameter $m$ and the interaction strength $U$ for $N \times 4$ cylinders with $N=4,8$. 

For increasing interaction strength $U$ and gap parameter $m\geq 2$, the sequence of phases is similar.  Starting with weak interactions, now there is a non-magnetic band insulating phase (green) which undergoes a transition to a paramagnetic topological phase T$_1$ (red) followed by a narrow region of stripey antiferromagnetic topological T$_2$ phase for intermediate coupling strengths.  Lastly, there is an antiferromagnetic Mott insulating phase in the strong coupling regime. On the other hand, for $0<m<2$  the non-magnetic band insulator at weak coupling is not observed. Here, with increasing interaction strength,  we find a paramagnetic topological insulating T$_1$ region, followed by a narrow region of stripey antiferromagnetic topological insulating T$_2$ phase, which turns into a robust antiferromagnetic Mott insulator for strong interactions. In Fig.~\ref{Fig: Magnetic and topological phase diagram}, the phase boundaries separate measured points in distinct phases. The overall distinction between topological and non-topological phases is consistent with a previous  DMFT study \cite{trauzettel13}. However, here the DMRG allows us to resolve the stripey AFM correlations in the topological phase near the boundary to the AFM Mott insulator.

To compute the phase diagram, we adhered to the convergence criteria outlined in our method section~\ref{Sec: Model and Method}.  The majority of our DMRG simulations (indicated as small solid/empty circles) are for $4\times 4$ cylinders, however we have verified the robustness of the observed phases for larger $8\times 4$ cylindrical systems (highlighted with with gray shading). Details of our finite size scaling analysis are included in Appendix~\ref{Sec: Finite-size scaling}. It is important to note that the solid circles indicate well-converged DMRG points, while the empty circles signify points that didn't converge effectively.

\begin{figure}[!b]
\centering
\includegraphics[width=3.4in,height=3.4in]{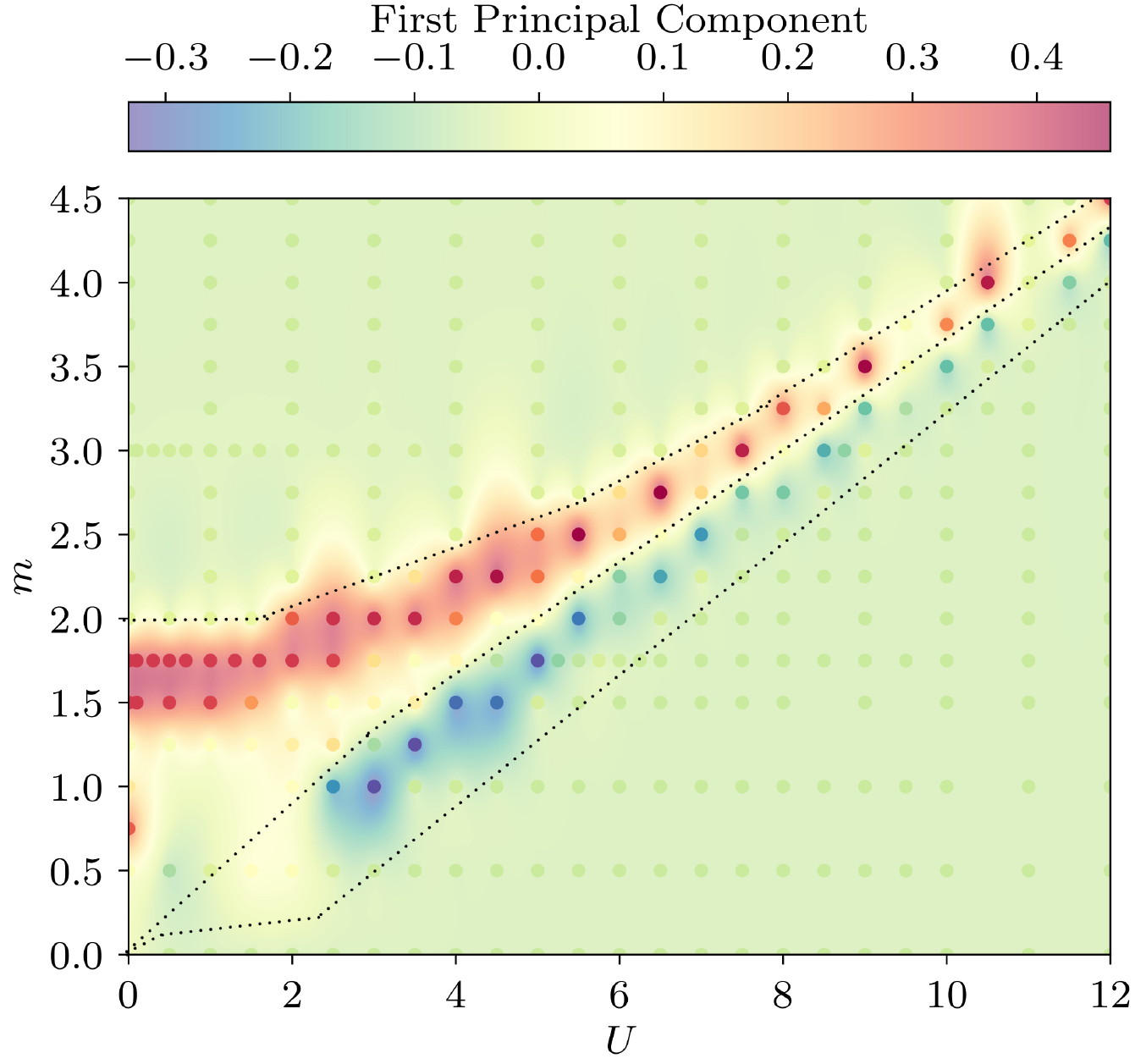}
\caption{Unsupervised machine learning phase diagram ($m$ vs $U$) of the interacting BHZ model on a $N \times 4$ cylinder colored by the value of the first principal component obtained from principal component analysis of the edge electronic charge densities. The green regions are comprised of the band and Mott insulating phases while the region with the red and blue points correlates with the topological phases in Fig.~\ref{Fig: Magnetic and topological phase diagram}. Radial basis function interpolation is used to extrapolate the remainder of the diagram.  Dotted lines are the same phase boundaries inferred from Fig.~\ref{Fig: Magnetic and topological phase diagram}. \label{Fig: PCA Phase Diagram}}
\end{figure}

\subsubsection{Verification of the phase diagram using unsupervised machine learning} 
As our analysis of topological properties was made without direct access to the energy spectrum, we further verify the phase diagram in Fig.~\ref{Fig: Magnetic and topological phase diagram} through an unsupervised machine learning approach. We constructed a data set from each point on the phase diagram, where we assign a concatenated array of the edge electronic charge density $\langle \tilde{n}_{i,\alpha}\rangle$ at each cell index for both the $s$ and $p$ orbitals. Principal component analysis (PCA) \cite{wang16,wetzel17,wang17,acevedo21} can be used to reduce the dimensionality of the data by identifying the mutually orthogonal directions along which the data varies the most through a linear combination of the original coordinates. We use the scikit-learn package in Python \cite{scikitlearn} to identify the direction of variation with the most variability. Figure~\ref{Fig: PCA Phase Diagram} depicts each point on the phase diagram colored according to the value of its first principal component, or position along the primary axis. Radial Basis Function interpolation is used to extrapolate the remaining segments of the phase diagram. Our analysis revealed that, while the band and Mott insulating phases exhibit similar alignment along the axis, the first principal component is clearly able to identify different signatures of the two topological phases, corresponding to a strong positive or negative signal. The use of spatially resolved raw orbital density data as input strongly supports their correlation with topological properties.

\section{Conclusion \label{Sec: Conclusion}}
This paper has addressed the challenge of incorporating electron-electron interactions into models of topological insulators.  In particular we have proposed and carried out  a simulation and analysis method to understand the effects of electronic correlations on the Bernevig-Hughes-Zhang model on a $N\times 4$ cylinder, using a numerically exact real-space density matrix renormalization group (DMRG) algorithm. By combining conventional magnetic order parameters and response functions with a real-space investigation of orbitally resolved densities, we describe the interplay between topological order and pronounced interaction effects emerging at the sample boundary. The observed signature of topological order, manifest in the electronic orbital polarization near the edge, opens a route for the study of strongly interacting topological systems via DMRG.  This analysis is supplemented via an unsupervised machine learning approach considering unlabelled spatial orbital occupancies as features. 

Our approach, which includes the full multi-orbital Hubbard interaction term, unveils a rich magnetic and topological phase diagram as a function of gap parameter $(m)$ and interaction strength $(U)$. At half-filling, our phase diagram reaffirms the existence of various magnetic and topological phases under the influence of interactions, in agreement with prior DMFT studies \cite{trauzettel13}, but also provides evidence for the presence of a more subtle antiferromagnetically ordered topological insulator \cite{topAFM1,topAFM2,topAFM3}.


While understanding the influence of interactions in topological matter remains a considerable challenge, the DMRG framework presented here may enable further theoretical and experimental exploration of strongly correlated topological insulators.

\section*{Data Availability}
All data, code, and analysis scripts that support the findings of this study can be found online \cite{codeRahul}.

\section*{Code Availability}
The \texttt{DMRG++} code used in this study is available at \href{https://g1257.github.io/dmrgPlusPlus/}{g1257.github.io/dmrgPlusPlus/} \cite{dmrgpp}.

\section*{Acknowledgments}
R.~S., H.~R., and A.~D. acknowledge support from the U.~S. Department of Energy, Office of Science, Office of Basic Energy Sciences,under Award No. DE-SC0022311. B.~R. acknowledges support from the German Research Foundation under grant RO 2247/11-1 and the hospitality of the University of Tennessee, where a portion of this work was performed. The U.S. Department of Energy, Office of Science, National Quantum Information Science Research Centers, Quantum Science Center has supported G.~A., who contributed to the DMRG aspects in this paper. R.~S. acknowledges the office of information technology (OIT) at the University of Tennessee for providing additional computational resources to carry out this project. The authors would like to thank F.~Heidrich-Meisner, H.~Barghathi, P.~Laurell and E.~Dagotto for useful discussions.

\appendix

\section{Single cell picture of the interacting BHZ model}
The effective Hamiltonian of an interacting unit cell with two orbitals ($s$ and $p$) is expressed as:
\begin{align}
H_{eff} &= m (n_{\vphantom{i^i}s} - n_{\vphantom{i^i}p}) + U \left( n_{\vphantom{i^i}s,\uparrow}n_{\vphantom{i^i}s,\downarrow} + n_{\vphantom{i^i}p,\uparrow}n_{\vphantom{i^i}p,\downarrow} \right) \nonumber \\
& + \left(U'-\frac{J_{H}}{2}\right)n_{\vphantom{i^i}s}n_{\vphantom{i^i}p} - 2J_{H} \left(\mathbf{S}_{\vphantom{i^i}s}\cdot\mathbf{S}_{\vphantom{i^i}p}\right) \nonumber \\
& + J_{H}\left( P_{s}^{\dagger}P_{p} + h.c \right).  
\label{eq:singleCell}
\end{align}
This single cell picture of the interacting BHZ model can be analyzed to obtain an intuition and understanding of the origin of the magnetic phases present in the interacting phase diagram in Fig.~\ref{Fig: Magnetic and topological phase diagram}.  While Eq.~\eqref{eq:singleCell} does not have any hopping connections, nor does it have any reference point to identify a paramagnetic or antiferromagnetic phase; the Hamiltonian does still provide relevant information about the roles played by individual terms. 

\begin{figure}[!b]
\centering
\includegraphics[scale=1.1]{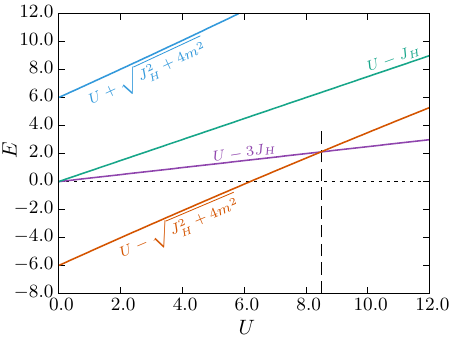}
\caption{Eigenvalues of an interacting unit cell versus interaction strength $U$ for $m=3.0$. The red and cyan lines represent the two band insulating states $\ket{B_1}$ and $\ket{B_2}$, respectively. The purple and green lines depict the spin triplet and singlet states, respectively. The plot shows the transition of the ground-state from a non-magnetic band-insulating state to a magnetic spin-triplet state at $U\simeq 8.5$ also indicated with a vertical dashed line. \label{Fig: Single cell picture m=3.0}}
\end{figure}

At half-filling, one can write the matrix form of this effective Hamiltonian using the following spin-basis $\lvert s,p\rangle := \{\lvert\uparrow,\downarrow\rangle, \lvert\downarrow,\uparrow\rangle, \lvert\uparrow\downarrow ,0\rangle, \lvert 0,\uparrow\downarrow\rangle \}$ as;
\begin{equation}
(H_{eff})=\begin{pmatrix}
U' & -J_H & 0 & 0 \\
-J_H & U' & 0 & 0 \\
0 & 0 & U+2m & J_H \\
0 & 0 & J_H & U-2m 
\end{pmatrix} . \label{Eqn: Single site Ham}
\end{equation}

Now let us understand the components of the Hamiltonian individually. The gap term alone ($U=J_H=0$) favors the $(n_s,n_p)=(0,2)$ configuration and thus dominates in the weakly interacting regime, whereas the Hubbard term by itself favors the $(1,1)$ configuration and takes over for strong interactions. It is then the competition that occurs between the $U'$ which explicitly pushes the system towards a mixed band insulator of $(0,2)$ and $(2,0)$ configuration, and the Hund's coupling and pair hopping term which forces the system towards the $(1,1)$ configuration, that governs the magnetic transitions in the intermediate interacting regime \cite{trauzettel13}.

The eigenvalues and corresponding eigenvectors of the matrix Hamiltonian in \eqref{Eqn: Single site Ham} are:
\begin{equation}
\begin{rcases*}
E_{B_1} = U-\sqrt{J_H^2 + 4m^2}, & $\lvert B_1\rangle = \gamma \lvert\uparrow\downarrow ,0\rangle + \delta \lvert 0,\uparrow\downarrow\rangle$\\
E_{B_2} = U+\sqrt{J_H^2 + 4m^2}, & $\lvert B_2\rangle = \delta \lvert\uparrow\downarrow ,0\rangle - \gamma \lvert 0,\uparrow\downarrow\rangle$\\
E_{S}  = U-J_H, &  $\lvert S\rangle = \frac{1}{\sqrt{2}}\left[\lvert\downarrow,\uparrow\rangle - \lvert\uparrow,\downarrow\rangle \right]$\\
E_{T}  = U-3J_H, & $\lvert T\rangle = \frac{1}{\sqrt{2}}\left[\lvert\downarrow,\uparrow\rangle +  \lvert\uparrow,\downarrow\rangle \right]$ \label{Eqn: S=3/2 Energies}
\end{rcases*},%
\end{equation}
where $\lvert B_1\rangle$ and $\lvert B_2\rangle$ are two band insulating states formed by the linear combinations of $(n_s,n_p)=(0,2)$ and $(2,0)$ configurations, with coefficients $\gamma=-J_H$ and $\delta=2m+\sqrt{J_H^2 + 4m^2}$. The states $\lvert S\rangle$ and $\lvert T\rangle$ are the spin singlet and triplet states with $S^z_{tot}=0$, formed by the $(1,1)$ configurations. Other than the triplet state all three states have zero magnetic moments and thus are non-magnetic.

\begin{figure}[!t]
\centering
\includegraphics[scale=1.1]{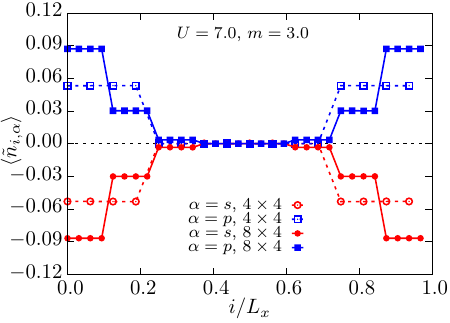}
\caption{Edge electronic charge density vs. the relative distance along the cylinder ($i/L_x$) for 8$\times$4 (filled) and 4$\times$4 (open symbols) systems using DMRG at $U=7.0$ and $m=3.0$.  Parameters are chosen such that the system lies within the paramagnetic T$_1$ phase. A finite orbital occupancy at the edge persists for the largest systems measured. \label{Fig: 4x4 and 8x4 comparison at U=7.0, m=3.0}}
\end{figure}

In Fig.~\ref{Fig: Single cell picture m=3.0}, we plot the above four eigenvalues for $m=3.0$. The plot clearly demonstrates that, from weaker to intermediate values of $U$ the ground-state lies in the non-magnetic band insulator phase in form of the $\lvert B_1\rangle$ state, and as it reaches the intermediate interacting regime it transitions to the magnetic triplet state $\lvert T\rangle$ which continues to the strongly interacting regime. The transition from the $\lvert B_1\rangle$ state to $\lvert T\rangle$ state occurs at $U=2\sqrt{2}m$, for $m=3.0$ it happens at $U=8.485(\simeq 8.5)$. Moreover, this transition point is consistently close to our DMRG phase boundary between the paramagnetic and antiferromagnetic phase in Fig.~\ref{Fig: Magnetic and topological phase diagram}.

\begin{figure}[!t]
\centering
\includegraphics[scale=1.25]{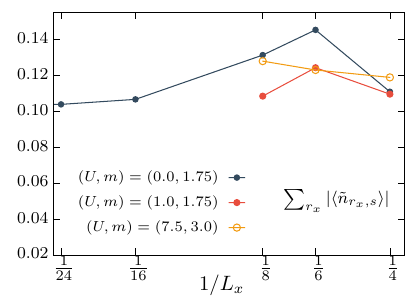}
\caption{Finite size scaling analysis of the edge density for $(U,m)=(0.0,1.75)$ (system sizes up to $24 \times 4$),  $(U,m)=(1.0,1.75)$ (systems up to $8 \times 4$), and $(U,m)=(7.5,3.0)$ (systems up to $8 \times 4$) all of which lies in the paramagnetic T$_1$ topological phase. The plot is obtained by taking the average area under curve of the edge electronic density, $\langle \tilde{n}_{r_x,s}\rangle=\sum_{r_y}\langle \tilde{n}_{r_x,r_y,s}\rangle/L_y$ and depicts robustness of the edge signals of orbital-$s$ as we move towards larger system sizes. \label{Fig: Finite size scaling for PM T1 phase}}
\end{figure}

\section{Finite-size scaling \label{Sec: Finite-size scaling}}

To explore the robustness of DMRG results on finite cylinders, we conducted a finite size scaling analysis of the observed topological phases. Figure~\ref{Fig: 4x4 and 8x4 comparison at U=7.0, m=3.0} presents the edge electronic charge density for 8$\times$4 and 4$\times$4 cylinders at $U=7.0$ and $m=3.0$, which correspond to the paramagnetic T$_1$ phase.  The results clearly show that the edge density signal in the 8$\times$4 cylinder are consistent and more robust as compared to the 4$\times$4 case. This consistency establishes the reliability of our topological marker as well as the separation between edge and bulk properties.

\begin{figure}[!h]
\centering
\hspace*{-0.5cm}\includegraphics[scale=0.87]{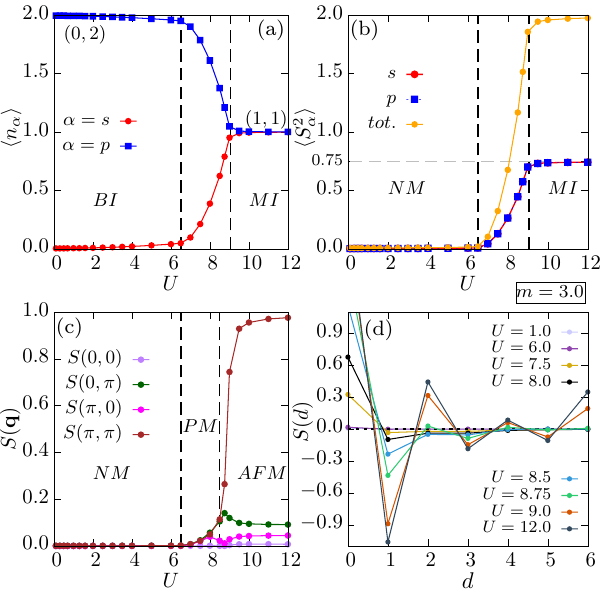}
\caption{Charge and magnetic properties of the interacting BHZ model with $m=3.0$, $U> 0$ on a $4\times 4$ cylinder. (a) Average orbital occupation vs. $U$ for both $s$ and $p$-orbitals, revealing a change in orbital occupation from a $(0,2)$ BI configuration to $(1,1)$ MI configuration with increasing interaction strength. (b) Average magnetic moment of a orbitals and of the entire cell are plotted versus $U$, showcasing a magnetic transition from a NM phase to a magnetic phase at $U=6.5$. (c) The spin structure factor plot demonstrating the transition from NM to PM to AFM phase as we increase $U$. (d) Real-space spin-spin correlations vs. the scalar distance between spins. These findings are consistent with our results at the same parameters for an $8\times 4$ system shown in Fig.~\ref{Fig: Magnetic properties m=3.0, U!=0, 8x4}.} \label{Fig: Magnetic properties m=3.0, U!=0, 4x4}
\end{figure}

To obtain more systematic information on finite size effects for the edge densities, we calculated $\langle \tilde{n}_{r_x,s}\rangle=\sum_{r_y}\langle \tilde{n}_{r_x,r_y,s}\rangle/L_y$, for three different set of parameters $(U,m)=(0.0, 1.75), (1.0, 1.75), (7.5, 3.0)$ from the PM T$_1$ phase with the results shown in Fig.~\ref{Fig: Finite size scaling for PM T1 phase}.  The plot illustrates that the signal at $(U,m)=(0.0,1.75)$ saturates to a finite value as we increase the size of the cylinder along the $x$-axis. More importantly, we observe a similar behavior in the presence of interactions for both $(U,m)=(1.0,1.75)$ and $(U,m)=(7.5,3.0)$ where DMRG calculations are computationally costly.

\section{$4 \times 4$ Results at $m=3.0$ \label{Sec: 4x4 Results at m=3.0}}
For the case of $m=3.0$, we have performed DMRG simulations for both $8\times 4$ and $4\times 4$ cylinders over all interaction strengths. We observed that in both cases, the magnetic and topological properties are remarkably consistent, indicating the reliability of our DMRG determined phase diagram

This can be clearly illustrated through Fig.~\ref{Fig: Magnetic properties m=3.0, U!=0, 4x4} and Fig.~\ref{Fig: Magnetic properties m=3.0, U!=0, 8x4} which depict the charge and magnetic properties of a $4\times 4$ and $8\times 4$ cylinders, respectively, at $m=3.0$. In both cases, the presence of different magnetic phases, and their ordering and transitions as we increase $U$, are consistent with each other. Similarly for the topological properties, a side-by-side comparison between Fig.~\ref{Fig: Topological properties m=3.0, U!=0, 4x4}(b) and Fig.~\ref{Fig: Topological properties m=3.0, U!=0, 8x4}(b) confirms the persistence and robustness of the T$_1$ and T$_2$ topological phases, across both cylinder sizes, within the framework of our DMRG calculations for the interacting BHZ model.

\begin{figure}[t]
\centering
\hspace*{-0.35cm}\includegraphics[scale=0.9]{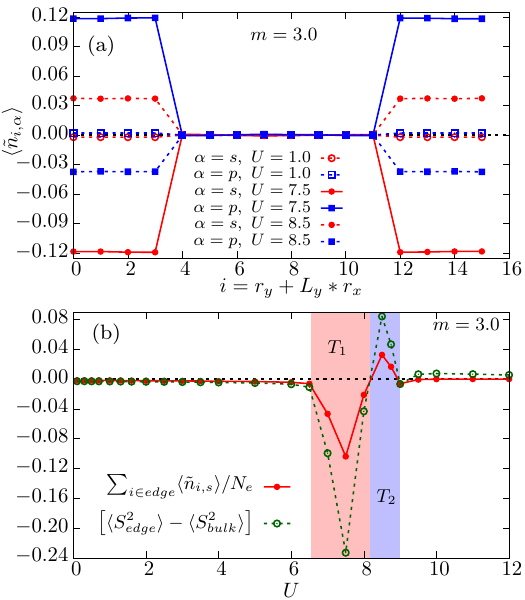}
\caption{Topological properties of the interacting BHZ model at $m=3.0$ for a $4\times 4$ cylinder. (a) The plot of edge electronic charge density versus the cell index $i$ for different values of $U=1.0,~7.5,~8.5$. The plot shows that the parameter point $U=1.0$ is non-topological or trivial, whereas the parameter points $U=7.5$ and $U=8.5$ corresponds to the T$_1$ and T$_2$ topological phases, respectively. (b) Average edge density for orbital-$s$ and the difference of edge and bulk magnetic moments vs. $U$, depicting a change in polarity of the edge density as distinguishing the T$_1$ and T$_2$ topological phases. The findings are consistent with our topological results at the same parameter for $8\times 4$ system in Fig.\ref{Fig: Topological properties m=3.0, U!=0, 8x4}.} \label{Fig: Topological properties m=3.0, U!=0, 4x4}
\end{figure}

\begin{figure}[t]
\centering
\includegraphics[scale=1]{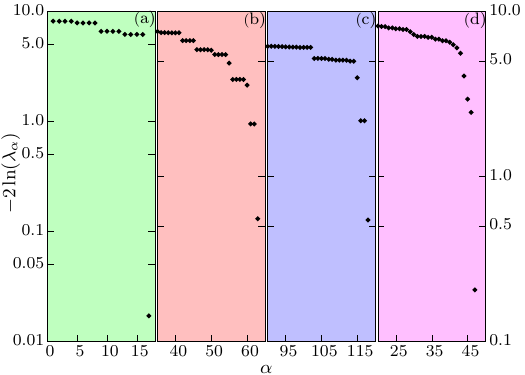}
\caption{The low-level entanglement spectrum on an $8\times 4$ cylinder at a fixed $m=3.0$ is illustrated for different values of $U$, corresponding to various phases in our study. In panel (a), we present the entanglement spectrum for $U=1.0$ for the band insulator, (b) $U=7.0$ from the paramagnetic topological T$_1$ (Paramagnetic T$_1$) phase, (c) displays the entanglement spectrum for $U=8.5$ from the SAFM T$_2$ (Stripey-Antiferromagnetic T$_2$) phase, and (d) depicts the entanglement spectrum  for $U=11.0$ from the AFM MI (Antiferromagnetic Mott Insulator) region. The $y$-axis on the left pertains to panel (a), while the right $y$-axis applies to panels (b), (c), and (d). The plot highlights the presence of degenerate density matrix eigenvalues in the first excited state of the spectrum: 4-fold degenerate in the BI phase, 2-fold in both the T$_1$ and T$_2$ phases, with the MI phase showing non-degeneracy. } \label{Fig: Entanglement spectra at m=3.0}
\end{figure}

\section{Entanglement spectrum}
The entanglement spectrum of the system can be obtained from the Schmidt decomposition of the many-body ground state $\ket{\Psi} = \sum_{\alpha}\lambda_{\alpha}\ket{L,\alpha}\otimes\ket{R,\alpha}$, where $\ket{L,\alpha} (\ket{R,\alpha})$ corresponds to the left(right) partitions of the system and $\lambda^{2}_{\alpha}$ are the eigenvalues of the spatially reduced density matrix $\rho_L= \text{Tr}_{R}\ket{\Psi}\bra{\Psi}$ of the partition \cite{haldane08}. Analyzing the eigenvalues can facilitate distinguishing between topological and non-topological phases, as demonstrated previously in 1D systems \cite{pollman10}. However, it is not a universal indicator for characterizing distinct phases \cite{sondhi14}. 
Figure~\ref{Fig: Entanglement spectra at m=3.0} shows a plot of the low-level entanglement spectrum with $m=3.0$ on an $8\times 4$ cylinder for various values of $U$, corresponding to different phases. The entanglement levels are obtained by diagonalizing the reduced density matrix for a center cut dividing the cylinder in two, where the cut is acting like an edge in the $y$-direction. In the band insulator spectrum seen in panel (a) for $U=1.0$, we observe a single low lying eigenvalue corresponding to the entanglement in the ground state separated by a large gap to several 4-fold degenerate excited states.  In case of the PM T$_1$ and SAFM T$_2$ phase, which are plotted for $U=7.0$ and $U=8.5$ as panels (b) and (c) we observe a doubly degenerate excited level corresponding to the helical nature of edge states.  In panel (d), the emergence of long range antiferrogmagnetic order breaks time reversal symmetry and leads to a splitting that is seen in the non-degenerate excited entanglement levels.



\FloatBarrier

\end{document}